# Systematic comprehension the metal phase of $Pr_{0.7}(Ca_xSr_{1-x})_{0.3}MnO_3$ via temperature and magnetic induction

Liu Changshi

Nan Hu College, Jiaxing University, Zhejiang, 314001, P. R. China， E-mail: lcswl@mail.zjxu.edu.cn

## Abstract

Temperature, magnetic induction and substitution dependent resistivity are a crucial factor in determining the physical properties of magneto-resistive materials. The first objective of this work was to find out an applicable method of using temperature to predict the resistivity of $Pr_{0.7}(Ca_xSr_{1-x})_{0.3}MnO_3$ in the metal phase within the transition area. Based on non-linear curve fitting, a typical numerical method is used to quantitatively analyze the temperature-dependent resistivity within temperatures lower than the metal-insulator transition temperature ($T_p$). The simulations agree very well with the observed curves (resistivity versus temperature). The second objective of this work is to search for the applicable method to link magneto-resistivity to magnetic induction, calculation on the principle of non-linear curve fitting, four typical numerical methods are highlighted because the magnetic induction-dependent magneto-resistivity are quantitatively analyzed by these methods. The observed effects of magnetic induction on the shift of temperature-resistive curve should be more essential in physics and are quantitatively discussed in magneto-resistive materials. Lastly, the influences of Ca substitution on magneto-resistivity are detected.

Keywords: Temperature; Resistivity; Magneto-resistivity; Simulation; Magnetic induction; Ca substitution.

1  Introduction

Since the discovery of colossal magneto-resistive effects in $Pr_{0.7}(Ca_xSr_{1-x})_{0.3}MnO_3$, the magnetic perovskite of the type $Pr_{0.7}(Ca_xSr_{1-x})_{0.3}MnO_3$ has attracted considerable research activity lately because of the rich physics involved as well as their potential use in device applications [1-4]. $Pr_{0.7}(Ca_xSr_{1-x})_{0.3}MnO_3$ are of great interest due to their high magneto-resistivity around room temperature. Physical properties, such as resistivity, $\rho$, the hopping energy of the small polaron, $E_{hop}$, the maximum resistivity, $\rho_{max}$, and the metal-insulator transition temperature, $T_p$, are of great interest in engineering processes as well as the progress of electromagnetic theories. Systematic research of those properties as a function of temperature, magnetic induction and substitution can give insight into the molecular structure of mixtures, provide information on the interaction between components, and are essential for designing and testing theoretical models of mixtures. It is believed that the phase transition of manganite materials can be predicted by analyzing the resistivity as a function of temperature. Spectrum of the temperature dependent resistivity of manganite material indicates its metal-insulator transitions during the temperature change and the transition order at a specific temperature. Manganite materials are ferromagnetic with a Curie temperature. The transition to magnetism is accompanied by a metal-insulator transition, the manganites being nonmetallic above and metallic below the Curie temperature. The metal-insulator transition also strongly depends on the applied magnetics induction, thus leading to a large "colossal" magneto-resistivity near the transition temperature. The maximum resistivity ($\rho_{max}$), $T_p$ and $E_{hop}$ represent different properties in separate FM and PM phase, respectively. In experiments, different methods used for changing the $\rho_{max}$, $T_p$ and $E_{hop}$ such as by changing element contents and by applying magnetic inductions, were applied to reveal correlations and competitions of couplings [5-11].

The most effective way to understand the phenomenon of magneto-resistivity is to find mathematical expressions by which the quantitative relationship between resistivity and temperature can be described. However, at present three models, i. e., Arrhenius law [12], polaron model [13], and variable-range hopping model [14], are used to fit temperature-dependent resistivity of the manganites in insulator state (T ≥ $T_p$), where, $T_p$ is the metal-insulator transition temperature. These three laws are very important in manganites research because they very well describe the observed high-temperature variation in the insulator mechanism. However, it has to be kept in mind that the pure Arrhenius law might be a very crude approximation [15]. The polaron model is valid only for temperatures higher than half of the Debye temperature $\Theta_D$ [15]. When the carriers are localized by random potential fluctuation, Mott`s variable-range hopping model is effective [15]. Although the above three laws can give an acceptable explanation for the relationship between the resistivity and the temperature in the paramagnetic phase, there is still no clear conclusion of whether or not the resistivity can be predicted by temperature in the range T ≤ $T_p$ of metal state with only one equation. The interest in resistivity is growing in the context of manganite. Unfortunately, few studies address this subject. Resistivity is usually one of the important physical parameters characterizing a new compound. However, accurate calculations of resistivity are quite difficult from a theoretical perspective. Physical experience leads people to believe that specification function on experimental results defines a unique physical problem，even if the specification function is empirical functional forms. Hence, in this work, a mathematical function was first chosen to predict resistivity by temperature which is lower than $T_p$ in the metal phase is as close as practically possible to $Pr_{0.7}(Ca_xSr_{1-x})_{0.3}MnO_3$ so that a more incisive probe on sorting out the results of the spectrum of temperature-resistivity can be provided. Based on the comparison between the shapes of experimental curves and the shape of the SRichards2 function, 40 resistivity curves as a function of temperature below $T_p$ are simulated in the following sections. Second, in order to forecast key physical parameters of manganites, the formula for comprehending the effects of both the magnetic induction and Ca content on the temperature-resistivity curve is proposed from data fitting of computer simulations. Satisfactory agreement is reached between the measured temperature-resistivity curve and the modeled one. Therefore, the proposed approach allows estimation of the shifts of the temperature-resistivity curve for individual manganite materials.

2. Simulation

The magneto-resistivity of manganite materials is determined by several factors such as composition, x, temperature, T, the magnetic induction, B, grain size and so on. The curves for temperature dependence of resistivity of $Pr_{0.7}(Ca_xSr_{1-x})_{0.3}MnO_3$ in several magnetic inductions in Fig. 1 are taken directly from published papers [16].

There is one peak in each temperature-resistivity curves in Fig. 1. The temperature dependence of the measured resistivity (Fig. 1) is typical for magneto-resistive materials, exhibiting a two-phase behavior with a more conductive (metallic) and a less conductive (insulating) region separated by a pronounced maximum at the peak temperature $T_p$. The peak temperature is also known as the metal-insulator transition temperature. This characteristic shows that resistivity increases slowly in the range from 0 to $T_p$ and drops down rapidly when the temperature is higher than $T_p$. Therefore, every curve is a cusp. The insulator mechanism of manganites at high temperature ($T \geqslant T_p$) has been explained by the simple Arrhenius law, the polaron model, and the variable-range hopping model law in the past.

It can be concluded that the conductive patterns shown in Fig. 1 can be analyzed in detail with fitted parameters of the desired model. The tactful way to find the mathematical relationship between the resistivity and the temperature in the metal phase (T≤$T_p$) is to fit the curve. Such fitting would help people to find which model is the best suited for the present system as well as the nature of interaction. There are two paths for curve fitting. One way is to directly apply the fitting function; another way is to create a mathematical model when there is no appropriate fitting function available. After checking the fitting function, it is found that the first strategy is effective enough in this work and the SRichards2 function offers such an opportunity. The applicable SRichards2 function is expressed by Eq. (1)

$$\rho(x_0, B_0, T) = \rho(T) = \frac{\rho_{max}}{\sqrt[k]{1 + k \cdot \exp(-(T - T_c)/T_0)}} \quad (T \leqslant T_p) \quad (1)$$

Where $\rho_{max}$ is the maximum resistivity, $T_c$ is the critical temperature, if temperature is below $T_c$, exponential function $\exp(-(T - T_c)/T_0)$ will be greater than one, while, when the temperature is equal to or higher than $T_c$, exponential function $\exp(-(T - T_c)/T_0)$ will be smaller than one. $T_0$ is probable superconductor temperature and k is adjusting factor in scope. $\rho_{max}$, $T_c$, $T_0$ and k will be optimized. If the SRichards2 function

(1) is available for predicting resistivity, $\rho$, of magneto-resistive materials at temperatures below $T_p$, the maximum value of $\rho$ is given by $\rho_{max}$. Hence, the physical significance of parameter $\rho_{max}$ is the resistivity of manganite materials at transition temperature $T_p$.

This paper analyzes the resistive curves in metal state of $Pr_{0.7}(Ca_xSr_{1-x})_{0.3}MnO_3$ at different magnetic inductions according to the above approach (1). Then through experimental raw data employed in the method of the regression analysis, the experimental spectra for $Pr_{0.7}(Ca_xSr_{1-x})_{0.3}MnO_3$ in various magnetic inductions have been simulated using the component spectra shown in Fig. 1. Optimized parameters employed to simulate the component spectra of $Pr_{0.7}(Ca_xSr_{1-x})_{0.3}MnO_3$ are also listed in Table 1. Modeled verification is also carried out by a comparison of modeled resistivity obtained by numerical integration of resistivity with measured resistivity in Fig. 1. The comparison between the results of the chosen peaks and the best-fitted value of parameter $T_0$ by the SRichards2 function (1) indicates that parameter $T_0$ corresponds to the minimum value of $\rho$. As is shown in Table 1 and Fig. 1, it can be stated that a satisfactory agreement between the measured and the modeled $\rho$ is achieved by the functions given in Eq. (1). This good agreement implies that the theoretical values of the minimum and the maximum resistivity of manganite materials can be estimated in an easier way by the SRichards2 function simulation because it is effective in the metal state.

Resistivity of manganite materials is not only a function of temperature, but it is also a function of magnetic induction. If the mathematical relationships between magnetic induction and parameters $\rho_{max}$, $T_c$, $T_0$ and k can be obtained, it is possible for one to predict resistivity of $Pr_{0.7}(Ca_xSr_{1-x})_{0.3}MnO_3$, in other words, the resistivity of $Pr_{0.7}(Ca_xSr_{1-x})_{0.3}MnO_3$ in any magnetic induction and at some temperature lower than $T_p$ can be calculated by the method expressed in this paper even if experiment is not carried out. In order to obtain $\rho_{max}$ of $Pr_{0.7}(Ca_{0.5}Sr_{0.5})_{0.3}MnO_3$ as a parameter from the directly observed data; the model parameter of $\rho_{max}$ corresponding to the magnetic induction is plotted in Fig. 2. The non-linear curve fitting is one of the best ways to obtain the relationship between magnetic induction and $\rho_{max}$. Fig. 2 illustrates the plot used to determine $\rho_{max}$ by magnetic induction. Although magnetic induction dependence of $\rho_{max}$ is complex in Fig, 2, the equation to link magnetic induction to $\rho_{max}$ is ExpDecay1 function and which is written as following

$$\rho_{max}(B) = \rho_{max\,0} + \rho_{max\,1} \exp(-B/B_0) \tag{2}$$

where B is magnetic induction. Using function (2), the values of $\rho_{max0}$, $\rho_{max1}$ and $B_0$ are shown in Fig. 3.

The influences of magnetic induction on the parameter of k should be addressed. There is some hope to undertake the quantitative relationship between magnetic induction and k, Fig. 2 exhibit the results of magnetic induction dependence of k for $Pr_{0.7}(Ca_xSr_{1-x})_{0.3}MnO_3$. The best-fitted results show that linear equation successfully gives a quantitative relationship between k and magnetic induction via non-linear curve fitting, and linear function is written as

$$k(B) = k_1 + k_2 \cdot B \tag{3}$$

Each data of $k_1$ and $k_2$ for $Pr_{0.7}(Ca_xSr_{1-x})_{0.3}MnO_3$ is listed in Fig. 3.

In order to calculate parameter $T_c$ from the magnetic induction, the $T_c$ of $Pr_{0.7}(Ca_xSr_{1-x})_{0.3}MnO_3$ as a function of magnetic induction is plotted in Fig. 2. The non-linear curve fitting is the best method for obtaining the relationship between magnetic induction and $T_c$. the $T_c$ data of the present investigation are found to fit well with the linear function

$$T_c(B) = T_{c1} + T_{c2} \cdot B \tag{4}$$

The values of $T_{c1}$ and $T_{c2}$ for $Pr_{0.7}(Ca_xSr_{1-x})_{0.3}MnO_3$ are given via

Fig. 3.

There is some hope to reveal the quantitative relationship between magnetic induction and the value of parameter $T_0$. Fig. 2 illustrates the results of magnetic induction dependence of $T_0$ for $Pr_{0.7}(Ca_xSr_{1-x})_{0.3}MnO_3$. The best-fitted results show that the linear equation successfully gives a quantitative relationship between $T_0$ and magnetic induction via non-linear curve fitting and the linear function is written as

$$T_0(B) = T_{01} + T_{02} \cdot B \tag{5}$$

Every value of $T_{01}$ and $T_{02}$ for $Pr_{0.7}(Ca_xSr_{1-x})_{0.3}MnO_3$ is also shown in Fig. 3. Combining with equation (1), (2), (3), (4) and (5) function leads to the following expression for the resistivity of $Pr_{0.7}(Ca_xSr_{1-x})_{0.3}MnO_3$ in the metal phase.

$$\rho(x_0, B, T) = \rho(B, T) = \frac{\rho_{max\,0} + \rho_{max\,1} \cdot \exp(-B/B_0)}{\sqrt[k_1 + k_2 \cdot B]{1 + (k_1 + k_2 \cdot B) \cdot \exp(-(T - (T_{c1} + T_{C2} \cdot B))/(T_{01} + T_{02} \cdot B))}} \quad (T \leqslant T_P) \tag{6}$$

Therefore, it is good enough for one to predict resistivity of $Pr_{0.7}(Ca_xSr_{1-x})_{0.3}MnO_3$ from both temperature and magnetic induction.

The function (6) consistently explains the relation between magnetic induction and resistivity for $Pr_{0.7}(Ca_xSr_{1-x})_{0.3}MnO_3$, this may be due to the fact that delocalization of charge carriers are induced by magnetic induction some, which in turn might suppress the resistivity and also cause local ordering of the magnetic spins at the same speed. Due to this ordering, the ferromagnetic metallic (FMM) state may suppress the paramagnetic insulating (PMI) regime. As a result, the conduction electron ($e_g^1$) are completely polarized inside the magnetic domains and are easily transferred between the pairs of $Mn^{3+}$ ($t_{2g}^3 e_g^1 : S=2$) and $Mn^{4+}$ ($t_{2g}^3 e_g^{01} : S=3/2$) via oxygen.

In order to reveal correlations and competitions of couplings, the maximum resistivity and $T_c$ are changed by adjusting element contents in experiments. After detecting the effects of magnetic induction on resistivity of $Pr_{0.7}(Ca_xSr_{1-x})_{0.3}MnO_3$, it is natural to study the effects of Ca substitution concentration on metal phase of $Pr_{0.7}(Ca_xSr_{1-x})_{0.3}MnO_3$. As what shown in insensitive to Fig. 2 and 3, there is symmetry of chemical composition, the composition of symmetry is $Pr_{0.7}(Ca_5Sr_5)_{0.3}MnO_3$. The maximum $T_0$ and $\rho_{max}$ are belong to $Pr_{0.7}(Ca_5Sr_5)_{0.3}MnO_3$, while, the minim k and $T_c$ also are belong to $Pr_{0.7}(Ca_5Sr_5)_{0.3}MnO_3$. The influence of chemical composition on $T_{c1}$ is remarkable, but $\rho_{max0}$, $\rho_{max1}$, $B_0$, $k_1$, $k_2$, $T_{c2}$, $T_{01}$ and $T_{02}$ are insensitive to chemical composition. $\rho_{max0}$, $\rho_{max1}$, $B_0$, $k_1$, $k_2$, $T_{c2}$, $T_{01}$ and $T_{02}$ cannot be linked to chemical composition via mathematical method within the experimental range.

4 Conclusions

In this research, 40 resistivities of $Pr_{0.7}(Ca_xSr_{1-x})_{0.3}MnO_3$ in differential magnetic induction was calculated as a function of temperature in the metal state ($T \leq T_p$). A typical mathematical function was applied. Not only can the function describe the change trend of the electrical resistivity at temperature in the metal range, but also it predicts some characteristic indexes of manganite materials. The theoretical data simulated in this paper are consistent with experimental value, the good agreement between the measured and the simulated results of resistivity indicates that the SRichards2 function can accurately predict some aspects of manganite materials in the metal state. All the data of $\rho_{max}$, $T_c$, $T_0$ and k collected from $Pr_{0.7}(Ca_xSr_{1-x})_{0.3}MnO_3$ exhibit regular distributions. Therefore, the $\rho_{max}$, $T_c$, $T_0$ and k of $Pr_{0.7}(Ca_xSr_{1-x})_{0.3}MnO_3$ was calculated as a function of magnetic induction. According to the description of ferromagnetism in the double-exchange model, the observed and simulated relationship between the parameter in equation to link temperature and resistivity and magnetic induction should be more essential in physics and important in understanding the nature of magneto-resistive materials.

The influences of Ca concentration on magneto-resistivity are explained, the mathematical results for five kinds of relationship state that the magneto-resistivity properties of $Pr_{0.7}(Ca_xSr_{1-x})_{0.3}MnO_3$ can be controlled precisely. More importantly, mathematical results present a simple physics picture and useful knowledge for understanding the origin of Colossal Magnetoresistivity in such complex strongly correlated electronic materials of manganite. Data and data analysis provide the information would be further studies for comprehensive the doped manganite.

Acknowledgements

The author would like to thank Dr. Ma Yu-Bin from physics department of Peking University for providing the experimental convenience.

Table 1 Results and evaluations of simulation on temperature-dependent resistivity of $Pr_{0.7}(Ca_xSr_{1-x})_{0.3}MnO_3$ ($T \leq T_p$) in different magnetic induction.

| magnetic induction (T) | $Pr_{0.7}(Ca_xSr_{1-x})_{0.3}MnO_3$ | $\rho_{max}(\Omega cm)$ | $T_c$ (K) | k | $T_0$(K) |
|---|---|---|---|---|---|
| 0 | $Pr_{0.7}(Ca_{0.5}Sr_{0.5})_{0.3}MnO_3$ | 3.45 | 191.51 | 2.91 | 2.86 |
| 1 | | 2.69 | 197.14 | 3.1 | 3.45 |
| 2 | | 2.02 | 204.13 | 3.23 | 3.85 |
| 3 | | 1.56 | 210.43 | 3.33 | 4.35 |
| 4 | | 1.24 | 216.25 | 3.48 | 4.54 |
| 5 | | 1.01 | 221.69 | 3.70 | 5 |
| 6.5 | | .78 | 229.46 | 4.10 | 5.56 |
| 8 | | .63 | 236.31 | 4.26 | 5.99 |
| 4 | $Pr_{0.7}(Ca_{0.3}Sr_{0.7})_{0.3}MnO_3$ | .78 | 251.65 | 5.91 | 3.85 |
| 4 | $Pr_{0.7}(Ca_{0.4}Sr_{0.6})_{0.3}MnO_3$ | .88 | 240.63 | 4.82 | 4 |
| 4 | $Pr_{0.7}(Ca_{0.6}Sr_{0.4})_{0.3}MnO_3$ | .88 | 240.63 | 4.83 | 4 |
| 4 | $Pr_{0.7}(Ca_{0.7}Sr_{0.3})_{0.3}MnO_3$ | .78 | 251.66 | 5.96 | 3.85 |

Figure caption

Fig. 1 Comparisons between experimental and calculated temperature, *T*, dependence of the electrical resistivity for $Pr_{0.7}(Ca_xSr_{1-x})_{0.3}MnO_3$ measured in several magnetic inductions.

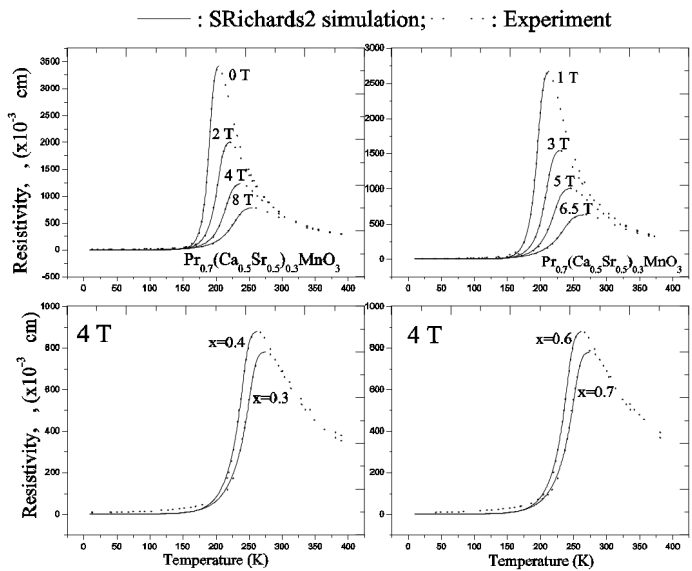

Fig. 2 Comparisons between optimized and fitted parameters $\rho_{max}$, k, $T_c$ and $T_0$ dependence of the magnetic inductions for $Pr_{0.7}(Ca_xSr_{1-x})_{0.3}MnO_3$.

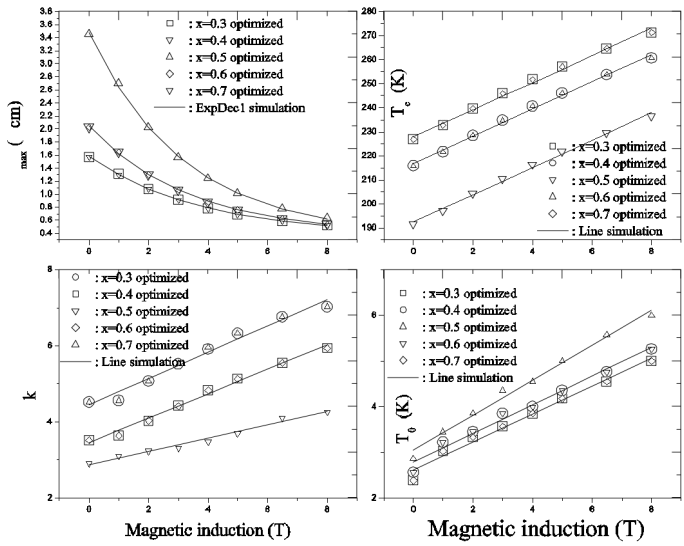

Fig. 3 Scatter plot of parameters $\rho_{max0}$, $\rho_{max1}$, $B_0$, $k_1$, $k_2$, $T_{c1}$, $T_{c2}$, $T_{01}$ and $T_{02}$ against the Ca concentration $Pr_{0.7}(Ca_xSr_{1-x})_{0.3}MnO_3$.

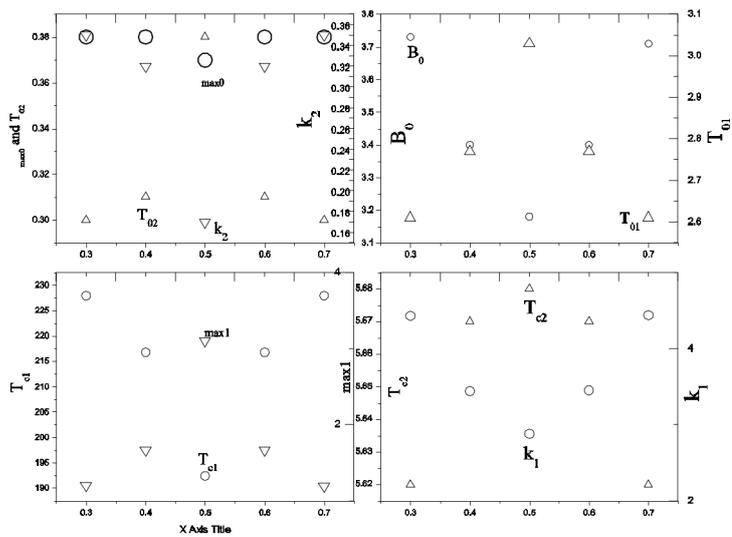